\documentclass[12pt]{iopart}
\usepackage[dvips]{graphicx}

\begin{document}

\def\be{\begin{equation}}
\def\ee{\end{equation}}
\def\ba{\begin{eqnarray}}
\def\ea{\end{eqnarray}}

\title[$k$--Body Embedded Gaussian Ensembles]{Review of the $k$--Body
Embedded Ensembles of Gaussian Random Matrices}

\author{L. Benet}
\address{Centro de Ciencias F\'{\i}sicas, U.N.A.M. (Unidad Morelos),
62210 Cuernavaca, Mor., M\'exico}
\author{H.A. Weidenm\"uller}
\address{Max--Planck--Institut f\"ur Kernphysik, 
D-69029 Heidelberg, Germany}

\begin{abstract}
The embedded ensembles were introduced by Mon and French as physically
more plausible stochastic models of many--body systems governed by
one--and two--body interactions than provided by standard
random--matrix theory. We review several approaches aimed at
determining the spectral density, the spectral fluctuation properties,
and the ergodic properties of these ensembles: moments methods,
numerical simulations, the replica trick, the eigenvector
decomposition of the matrix of second moments and supersymmetry, the
binary correlation approximation, and the study of correlations
between matrix elements.
\end{abstract}



\section{Introduction}
\label{int}

Canonical random--matrix theory (RMT) as introduced by
Wigner~\cite{wig51} considers ensembles of random matrices
classified~\cite{dys62} by their symmetries. In the asymptotic limit
of infinite matrix dimension, $N \to \infty$, this theory makes a
number of remarkable predictions. The average spectrum (the spectral
density) has the shape of a semicircle, the spectral fluctuations are
universal (i.e., are under very weak conditions independent of the
weight function used to define the ensemble, are parameter--free, and
dependent only on the symmetries of the ensemble), and the results are
ergodic (so that for most members of the ensemble the spectral average
of any observable coincides with the ensemble average of that same
observable). For a recent review of the field, we refer to
reference~\cite{guh98}.

Nevertheless, early applications of canonical RMT to nuclear spectra
encountered objections or, at least, questions. This is because in
RMT, the number of independent random variables (i.e., of uncorrelated
matrix elements) is of order $N^2$. Put differently, in RMT every
state in Hilbert space is connected to every other such state by a
matrix element which does not vanish and is assumed to be an
independent random variable. But with the success of the nuclear shell
model, it had become clear that nuclei are governed by one-- and
two--body interactions (the mean field and the residual two--body
interaction).  In a representation of Hilbert space spanned by Slater
determinants of single--particle wave functions (solutions of the
mean--field equations), the residual two--body interaction has
non--vanishing matrix elements only between determinants which differ
in the occupation numbers of not more than two single--particle
states, and most matrix elements of the residual two--body
interaction, therefore, vanish. Even if one is prepared to accept a
stochastic approach and considers the non--vanishing two--body matrix
elements of the residual interaction as uncorrelated random variables,
the ensuing stochastic model is very different from canonical RMT: The
number of independent random variables is generically small compared
to the dimension of a typical shell--model matrix. Why --- so the
question --- can canonical RMT then serve as a model for the
description of fluctuation properties of nuclear spectra?

French and Wong~\cite{fre70,fre71} and Bohigas and
Flores~\cite{boh71,boh71a} approached this question with the help of
numerical simulations. These authors replaced the actual two--body
matrix elements of the nuclear shell model by random variables (this,
in essence, defines the two--body random ensemble (TBRE)) and studied
the resulting spectral fluctuation properties. They concluded that the
fluctuation properties were consistent with predictions of canonical
RMT.

Because of the complexities of angular--momentum and spin coupling and
the exclusion principle, the TBRE is not amenable to analytical
investigations. Thus, the introduction of the $k$--body embedded
ensembles (EGE($k$)) by Mon and French~\cite{mon75} might be
considered an essential step towards an analytical understanding of
the numerical results obtained by French and Wong~\cite{fre70,fre71}
and Bohigas and Flores~\cite{boh71,boh71a} and, thereby, of the
stochastic properties of nuclear spectra. Since then, the question has
gained much wider significance: Canonical RMT has successfully been
used~\cite{guh98} in such diverse many--body systems as atoms,
molecules, and quantum dots. All of these systems share with nuclei
the feature that they are known to be governed essentially by
effective one-- and two--body forces. Thus, the embedded ensembles may
be viewed as the generic models for stochasticity in many--body
systems.

The embedded ensembles dispose of all the complexities of the
couplings of angular momentum and spin while retaining the symmetries
imposed by the exclusion principle. One considers $m$ Fermions in $l$
degenerate single--particle states interacting via a random $k$--body
interaction. The single--particle states carry no further quantum
numbers like spin or angular momentum. In contrast to shell--model
calculations (where the single--particle states are usually not
degenerate) degeneracy is here assumed in order to focus attention on
the results of a stochastic $k$--body interaction. The interaction is
not restricted to $k = 2$ (although this remains the physically most
interesting case). This is done in order to obtain an understanding of
the transition from the case $k = 2$ to the case $k = m$ which, as it
turns out, is equivalent to canonical RMT. Just as in RMT, the
embedded ensembles can be bestowed with unitary, orthogonal or
symplectic symmetry, and one distinguishes accordingly EGUE($k$),
EGOE($k$), and EGSE($k$). As in canonical RMT, one is interested in
universal properties. Therefore, one considers the limit of infinite
matrix dimension. This limit is attained by letting the number of
single--particle states go to infinity, $l \to \infty$. EGE($k$) is,
thus, a generic model for stochasticity in many--body systems governed
by $k$--body interactions. Needless to say, EGE($k$) can and has been
generalized to the case of Bosons.

Central questions in the theory of the embedded ensembles then are:
(i) What is the shape of the spectral density? (ii) What are the
spectral fluctuation properties? (iii) Are these properties universal?
(iv) Are the spectra ergodic?

In this Review, we present the status of the field in the light of
these questions. We start with a definition of EGE($k$). Then, we
order the material by the analytical or numerical technique which has
been used, rather than by the properties of EGE which have been
addressed.

\section{Definitions: The $k$--body Embedded Ensembles of Fermions
and Bosons}
\label{defi}

We define the $k$--body embedded ensembles for Fermions and for Bosons
for the cases of unitary and of orthogonal symmetry. We do not
consider symplectic symmetry throughout this paper.

We consider a set of $l$ degenerate single--particle states labelled
$| j \rangle$, with $j = 1, \ldots, l$. The associated creation and
annihilation operators are denoted by $a^{\dagger}_j$ and $a_j$ in the
case of Fermions and $b^{\dagger}_j$ and $b_j$ in the case of
Bosons. These operators obey the usual (anti)commutation relations.
Then, $| j \rangle = a^{\dagger}_j |0 \rangle$ or $| j \rangle =
b^{\dagger}_j |0 \rangle$, with $|0\rangle$ the vacuum state.

To define the $k$--body interaction, it is useful to introduce
operators which create a normalized state with $k \leq l$ Fermions
or $k$ Bosons from the vacuum. In the case of Fermions, these are
written as
\be
\psi^{\dagger}_{k,\alpha} = \psi^{\dagger}_{j_1,j_2,\ldots,j_k} =
\prod_{s = 1}^k a^{\dagger}_{j_s},
\label{1}
\ee 
where $j_1 <j_2 < \ldots < j_k$. The label $k$ defines the number
of single--particle creation operators, while $\alpha$ stands for the
set of $j$'s. The corresponding annihilation operators are given by
\be 
\psi_{k,\alpha} = (\psi^{\dagger}_{k,\alpha})^{\dagger} \ .
\label{2}
\ee
In the case of Bosons, the $k$--body operators have the form
\be
\chi^{\dagger}_{k,\alpha} = \chi^{\dagger}_{j_1,j_2,\ldots,j_k} =
N_{\alpha} \prod_{s = 1}^k b^{\dagger}_{j_s},
\label{5}
\ee 
where $j_1 \leq j_2 \leq \ldots \leq j_k$. Here, $N_{\alpha}$ is a
normalization factor which guarantees that the state
$\chi^{\dagger}_{k,\alpha} | 0 \rangle$ has norm unity. For every set
of equal indices $j$ containing $n$ elements occurring in $\alpha$,
$N_{\alpha}$ contains the factor $1/\sqrt{n!}$.

The random $k$--body interaction for Fermions is given by
\be
V_k = \sum_{\alpha \gamma} v_{k;\alpha \gamma} \psi^{\dagger}_{k,
\alpha} \psi_{k,\gamma} \ .
\label{3}
\ee
The coefficients $v_{k;\alpha \gamma}$ are random variables with a
Gaussian probability distribution, mean value zero, and a common
second moment. The $V_k$ form an ensemble of random operators. In
the case of unitary symmetry, the $v_{k;\alpha \gamma}$ are complex.
The only non--vanishing second moment is
\be
\overline{v_{k;\alpha \gamma} v_{k;\alpha' \gamma'}^*} = v^2 \
\delta_{\alpha \alpha'} \delta_{\gamma \gamma'} \ .
\label{4}
\ee
The bar denotes the ensemble average. The Kronecker delta's stand
for the string $\delta_{j_1 j'_1} \delta_{j_2 j'_2} \ldots$. In the
case of orthogonal symmetry, the $v_{k;\alpha \gamma}$ are real and
obey
\be
\overline{v_{k;\alpha \gamma} v_{k;\alpha' \gamma'}} = v^2 \ [
\delta_{\alpha \alpha'} \delta_{\gamma \gamma'} + \delta_{\alpha
\gamma'} \delta_{\gamma \alpha'} ] \ .
\label{6}
\ee
Following the usage in canonical RMT, we distinguish both cases by
writing $V_k(\beta)$ with $\beta = 1,2$ for orthogonal and unitary
symmetry, respectively. The $k$--body random interaction for Bosons
is defined in complete analogy to equation~(\ref{3}). The parameter $v^2$
sets the scale of the energy. As long as we consider only a single
EGE($k$), we can put $v^2 = 1$ without loss of generality.

We consider a Hilbert space obtained by distributing $m \leq l$
Fermions or $m$ Bosons over the single--particle states $| j
\rangle$. A complete set of basis states for Fermions (Bosons) is
given by $\psi^{\dagger}_{m, \alpha} | 0 \rangle$ (by
$\chi^{\dagger}_{m, \alpha} | 0 \rangle$, respectively). The
dimensions of these two spaces are
\ba
N = {l \choose m} \ {\rm for \ Fermions} \ , \nonumber \\
\nonumber \\
N_B = { l + m - 1 \choose m} \ {\rm for \ Bosons} \ . 
\label{7}
\ea
Here and in the sequel, we use the index $B$ for Bosons while we
typically work without such an index for the more frequent case of
Fermions. We take $m \geq k$ and consider the matrix of the $k$--body
interaction $V_k$ in that $m$--particle Hilbert space. The matrix
elements $\langle \mu | V_k | \nu \rangle$ have the form
\be
\langle \mu | V_k | \nu \rangle = \langle 0 | \psi_{m, \mu} V_k
\psi^{\dagger}_{m,\nu} | 0 \rangle \ ,
\label{7a}
\ee
and analogously for Bosons. Using equation~(\ref{3}) for $V_k$, it is
easily seen that for $m = k$, the matrices of the $k$--body
interaction $V_k$ coincide with the GUE or the GOE, depending on the
symmetry chosen, both for Fermions and for Bosons. This is not the
case for $m > k$. In this case, we speak of the $k$--body embedded
Gaussian ensembles (EGE($k$)) of random matrices in the $m$--particle
Hilbert space. We use the notation EGUE($k$) and EGOE($k$) for the
cases of unitary and orthogonal symmetry, respectively. Aside from the
(unitary or orthogonal) symmetry, the embedded ensembles are defined
in terms of the three parameters $l, m, k$. We recall that in
canonical RMT, the limit $N \rightarrow \infty$ of infinite matrix
dimension yields universal results. For the EGE($k$), we proceed
likewise by taking the limit $l \rightarrow \infty$ which implies the
limit of infinite matrix dimension. For Fermions, we have the obvious
constraint that $k \leq m \leq l$.  The limit $l \rightarrow \infty$
with fixed $k$ can be attained by keeping $m$ fixed, or by keeping the
ratio $m/l \leq 1$ fixed. Brody {\it et al.}~\cite{bro81} define the
dilute limit by taking $l \to \infty, m \to \infty, m/l \to 0$. For
Bosons, $m$ may be larger than $l$. Since EGE($k$) is constructed in
such a way that EGE($m$) coincides with canonical RMT for both
Fermions and Bosons, and since in the dilute limit the distinction
between Fermions and Bosons becomes irrelevant, we expect that the
spectral properties of EGE($k$) do not differ qualitatively for
Fermions and Bosons except for the case where the number of $m$ Bosons
is close to or exceeds $l$, the dimension of the underlying
single--particle space. This is indeed what is found.

Canonical RMT is invariant under unitary or orthogonal transformations
of Hilbert space. While this statement carries over trivially to the
$k$--body ensembles embedded into a Hilbert space spanned by $k$
particles, that invariance is lost for $m > k$. The calculation of
spectral properties of canonical RMT greatly benefits from such an
invariance property. Thus, the determination of the spectral
properties of the embedded ensembles is much more difficult than for
canonical RMT.

\section{Moments Methods}
\label{mom}

The first analytical approach to EGE($k$), developed by Mon and
French~\cite{mon75}, calculates moments of the matrix elements of
$V_k$ and related quantities. It has led to insights into the shape of
the EGE($k$) spectra (the spectral density), has led to the
development of statistical spectroscopy in nuclei, and has given a
limited answer to the question of ergodicity. These topics are
reviewed in references~\cite{bro81,kot01}. The approach has mainly
been used for Fermions in the context of the nuclear shell model,
although it can also be applied for Bosons as in the Interacting Boson
Model for nuclei.  We confine ourselves to Fermions.

In a Hilbert space with $m$ Fermions, the $p^{\rm th}$ moment of $V_k$
is defined as $M_p(m) = (1/N) \ {\rm tr} \overline{ [ (V_k)^p ] }$. It
is obvious that all odd moments vanish. The even moments can be
evaluated using Wick contraction of the $V_k$'s. The moment is then the
sum over all patterns of pairwise contractions. Assigning the same
letter to pairs of Wick--contracted $V_k$'s, the fourth moment is, for
instance, given by $M_4(m) = (1/N) \ {\rm tr} \ ( 2 A_k A_k B_k B_k +
A_k B_k A_k B_k )$ where we have used the cyclic invariance of the
trace. For $l \rightarrow \infty$, the low moments can be worked out
using combinatorial arguments~\cite{mon75}. Knowledge of the moments
yields information on the average spectral density
$\overline{\rho(E)}$.  This is because of the relation \be {\rm tr}
\overline{ ( V_k )^p } = \overline{ \sum_i ( E_i )^p } = \int {\rm d}E
\ \overline{ \rho(E) } \ E^p \ ,
\label{8}
\ee
where the $E_i$ are the eigenvalues of $V_k$.

Canonical RMT is obtained for $k = m$ and predicts the spectral
density to have semicircular shape. On the other hand, in the dilute
limit, all Wick--contracted pairs of $V_k$ become independent of each
other, the moment $M_{2p}(m)$ is given by $M_{2p}(m) = (2p - 1)!!
(M_{2}(m))^p$, and the spectral density is Gaussian. This shows that
for fixed $k$ and for $m$ monotonically increasing from $m = k$, the
spectral density undergoes a transition from the semicircular to the
Gaussian shape. For $k = 1$, the transition can be worked out easily
because the level density for $m > 1$ is the convolution of $m$
semicircles. Already for $m = 9$, the spectral shape is very close to
Gaussian. This argument fails for $k \geq 2$. In
reference~\cite{mon75}, the even moments $M_{2p}(m)$ are worked out
for $p = 1, 2, 3, 4$. The values obtained are consistent with the
gradual transition from semicircular to Gaussian shape of the spectral
density.

In nuclear theory, sums over all final states of the strength of some
transition operator ${\cal O}$ play an important role. Examples are
single--particle transfer strengths or the Gamow--Teller transition
strength. Such transition strength sums can be worked out, too, using
the moments method. The reason is that $K = {\cal O}^{\dagger} {\cal
O}$ can be expressed in terms of creation and annihilation operators,
and terms like ${\rm tr} [ \overline{(V_k)^p K} ]$ can again be
calculated using combinatorial arguments, especially in the dilute
limit. This method has been very successfully applied in the framework
of the nuclear shell model. In many cases, EGOE($k$) results obtained
in this way show remarkable agreement with results of detailed
shell--model calculations. We do not dwell on this important issue
because an extensive review has appeared quite recently~\cite{kot01}.

Using the moments method to calculate the spectral density or
transition strength sums, we always use the ensemble average. Is it
justified to compare such averages with properties of a given physical
system (as we do when comparing with shell--model calculations)? While
a given physical system obviously does not permit averaging over the
(physically non--existent and fictitious) ensemble, it does permit the
determination of spectral averages. For instance, by grouping
neighbouring levels into bins, we can determine the mean spectral
density. This poses the question: Are spectral averages (taken over a
specific system) and ensemble averages (over the EGE($k$)) identical?
This problem is referred to as the problem of ergodicity, in analogy
to the well--known ergodic problem in classical statistical mechanics
concerning the equality of phase--space average and the long--time
average over a single trajectory.

For ensemble averaging to be relevant for an individual system, the
property under study must vary little from member to member of the
ensemble. In the context of the moments method, the question of
ergodicity is answered by calculating the variances
\be
\fl
\Sigma^{(2)}_{p,q}(m) = \overline{ (1/N) {\rm tr} (V_k^p) \ 
(1/N) {\rm tr} (V_k^q)} - \overline{ (1/N) {\rm tr} (V_k^p)} \cdot
\overline{ (1/N) {\rm tr} (V_k^q)} .
\label{9}
\ee
For canonical random--matrix theory and in a normalization where the
individual moments are finite in the limit $N \rightarrow \infty$,
the variances $\Sigma^{(2)}_{p,q}$ vanish asymptotically: The
canonical ensembles are ergodic as far as the spectral density is 
concerned. For the embedded ensembles, the calculation of the
variances in equation~(\ref{9}) involves contractions between pairs of
operators located in different traces as well as in the same trace.
The treatment of the first kind of pairs requires combinatorial
techniques beyond the ones developed for the moments~\cite{mon75}.
Mon and French could show that for EGOE($k$), $\Sigma^{(2)}_{p,q}(m)$
vanishes asymptotically for $l \rightarrow \infty$. Thus, they could
demonstrate ergodicity for the spectral density. Their argument can
easily be extended to EGUE($k$).

Unfortunately, the moments method, while very useful for the spectral
density and for statistical spectroscopy, is only of limited use for
calculating spectral fluctuations. For canonical RMT, Mon and French
consider density fluctuations as excitations of the semicircle and use
this approach to study spectral GOE fluctuations. They succeed in
re--deriving the logarithmic dependence of the spectral stiffness on
the length of the energy interval. However, the method apparently does
not permit the study of short--scale fluctuations like the
nearest--neighbour spacing distribution. Mon and French apply the same
method to EGOE($k$) with $m \gg k$. They show that long--range
fluctuations of the spectrum are large. As a consequence, neither the
position of a given (numbered) eigenvalue nor the position of the
centre of the spectrum or the variance are ergodic. We return to this
point in Section~\ref{eig}. The short--range fluctuations, however,
which yield information about spectral statistics, pose difficulties.
To quote from reference~\cite{bro81}: ``...there is no real theory yet
for EGOE fluctuations, the gap being one of the most significant ones
in the entire subject.''

\section{Numerical Results}
\label{num}

Insight into the spectral fluctuation properties of the embedded
ensembles may be gained from numerical simulations. Such simulations
have been performed from the early days of the field. Here we review
separately the simulations for Fermionic and for Bosonic systems.

\subsection{Fermionic Systems}
\label{sumf}

Prior to the introduction of EGE($k$) by Mon and French~\cite{mon75},
numerical simulations used the two--body random ensemble (TBRE). This
ensemble was introduced in the context of nuclear physics where
standard shell--model techniques were available to calculate spectra
of given spin and parity (and isospin if relevant) in terms of a fixed
two--body interaction. The latter was replaced by an ensemble of
two--body matrix elements with a Gaussian distribution, zero mean
value, and a common second moment. The antisymmetrized (Fermionic)
random two--body matrix elements thus define the residual interaction
among $m$ particles distributed over $l$ degenerate single--particle
states. Inasmuch as nuclei are governed by one-- and two--body forces,
the TBRE is clearly a more suitable stochastic model than canonical
RMT. The complexities arising from the angular--momentum and spin
couplings, from correlations between many--body matrix elements
induced by the two--body character of the interaction, and from
Pauli's exclusion principle made it impossible to treat the TBRE
analytically. This led French and Wong~\cite{fre70,fre71} and Bohigas
and Flores~\cite{boh71,boh71a} to resort to numerical simulations
using matrices of rather small dimensions ($\le 61$). The main
question addressed in these simulations was: Are the spectral
fluctuation properties of the TBRE the same as those of the GOE?

The main results of these early works are: (i)~The spectral density of
the TBRE is Gaussian, in agreement with the results of shell--model
calculations with realistic interactions. (ii)~The spectral density
displays a transition from Gaussian to semicircular shape as the rank
of the interaction $k$ is increased. (iii)~Unfolding the spectrum of
each member of the ensemble separately yields good agreement of the
fluctuation properties of the TBRE with those predicted by canonical
RMT. (iv)~The fluctuations of the position of the first eigenvalue of
the TBRE are stronger than those for the GOE. (v)~The TBRE is neither
stationary nor ergodic. These results hold also for simplified
versions of the TBRE, which do not involve, or at least do so only
partially, the complexity of the angular momentum algebra.

Early results on the $p^{\rm th}$ nearest neighbour spacing
distribution displayed systematic deviations between the TBRE and the
GOE~\cite{boh72} while experimental data showed agreement with GOE
predictions. This seemed to suggest the existence of an effective
many--body interaction in nuclei. It was soon realized, however, that
the deviations were related to the method of analysis of the spectra
and caused by the non--ergodic properties of the
TBRE~\cite{wong72,fre73}. More precisely, the deviations were due to
the difference between ensemble averaging and spectral averaging.
Ensemble averaging uses a single staircase function (the integrated
level density) to unfold the spectra of all members of the ensemble.
This function is obtained by superposing the levels of all members of
the ensemble. Early numerical results~\cite{fre70,boh71} (and later
analytical studies~\cite{gerv72,mon75}) had shown that in the dilute
limit ($l \gg m\gg k$, $l \to \infty$, $m \to \infty$ with $m/l \to
0$), the mean level density is Gaussian. This spectral shape was
accordingly used for ensemble unfolding. On the other hand, spectral
averaging uses a separate staircase function for each member of the
ensemble. It was found that the deviations from GOE behaviour obtained
by ensemble averaging were related to the variance of the spacing
distribution; an analytical estimate of this effect, linking ensemble
and spectral average, was also obtained~\cite{fre73,boh74}. A first
systematic comparison of spectral and (a corrected) ensemble
averaging, using matrices of rather small dimensions, was carried out
in reference~\cite{boh75}. This comparison showed remarkable agreement
between the TBRE results and the GOE predictions. The agreement
invalidated the earlier conclusion that nuclear interactions are
mainly of many--body type, and has left open the question of the rank
of the effective nuclear interaction~\cite{boh74,bro81}.

Unfolding the spectrum is technically a subtle task. It can
considerably influence the results on spectral fluctuations.
Attention was therefore devoted to a consistent way of constructing
ensemble averages which yield correct spectral
fluctuations~\cite{bro81,lab90}. This problem has recently again
attracted attention~\cite{flo01,jack01}. In the dilute limit, the
spectral density is Gaussian. However, knowledge of this fact is not
sufficient for a proper unfolding of the spectra of the EGE($k$) and
of the TBRE as obtained in numerical simulations with {\it
finite--dimensional} matrices. This is because for such matrices, the
first and second moments of the spectrum are not
ergodic~(cf. Section~\ref{low}). Hence, these moments must be
normalized for every member of the ensemble. The sensitivity of the
spectral fluctuations to these corrections was addressed in
reference~\cite{flo01}, where either the first or the second moment,
or both, are corrected for every matrix of the ensemble. The results
show that, in the center of the spectrum, the nearest--neighbour
spacing distribution is not sensitive to such corrections. For the
two--point function, ensemble unfolding with the Gaussian mean level
density yields results that deviate clearly from predictions of
canonical RMT. A slight correction results from adjusting the width of
the distribution for every member of the ensemble. A drastic
improvement is obtained when the distribution for every member of the
ensemble is recentered.  Applying both corrections together yields
results which are hardly distinguishable from those of canonical
RMT. Once the low moments of the spectral distributions are adjusted
so as to coincide for all members of the ensemble, a polynomial fit
for the ensemble staircase function is usually implemented. This
procedure yields a function with which the spectra can be meaningfully
unfolded.

A method for spectral unfolding which is somewhat more complicated but
has a firm theoretical basis~\cite{mon75,bro81} for the embedded
ensembles, uses a Gram--Charlier expansion. Starting from the fact
that in the dilute limit the spectral density is Gaussian, one writes
the spectral density $\rho_{\rm s}(x)$ of an ensemble of matrices of
finite dimension in the form~\cite{bro81,lab90}
\begin{equation}
\label{num:eq1}
\rho_{\rm s} (x) = \rho_{\rm G}(x) \left [ 1 + \sum_{n \ge 3} 
  {c_n {\rm H}_n(x) \over n!} \right].
\end{equation}
Here, $x=(E-E_c)/\sigma$ is the recentered energy in units of its
standard deviation, $\rho_{\rm G}$ is the Gaussian distribution and
${\rm H}_n(x)$ is the Hermite polynomial of degree $n$. The Hermite
polynomials describe the long--wavelength oscillations of $\rho_{\rm
s} (x)$ (long in terms of the mean level spacing) about the asymptotic
value $\rho_{\rm G}(x)$. The coefficients $c_n$ are adjusted so as to
minimize $\Delta_{\rm RMS}$, the overall root--mean--square error of
the level--to--level deviations of the staircase function calculated
from equation~(\ref{num:eq1}) from the data. Explicit expressions for
$c_3$ and $c_4$ are given in reference~\cite{lab90}. One uses
equation~(\ref{num:eq1}) with some maximum value $n_0$ in the sum over
$n$. This value should be sufficient to eliminate any secular trend in
$\Delta_{\rm RMS}$ while preserving the fluctuations of the
spectrum. The value of $n_0$ is usually determined by calculating
$\Delta_{\rm RMS}$ as a function of $n_0$, and by choosing the
smallest value beyond which no significant corrections to $\Delta_{\rm
RMS}$ are found. For the EGE($k$) this method yields good agreement
with the spectral fluctuation properties predicted by canonical
RMT~\cite{bro81,lab90,kot01}.

The spectral fluctuation properties at the edge of the spectrum have
also received increased attention. These are interesting for a
comparison of TBRE and EGOE($k$) results with the positions of
low--lying nuclear states. Bohigas and Flores~\cite{boh71a} compared
the properties of the low--lying part of the spectrum of the TBRE and
of the GOE. They showed that the widths of the positions of individual
eigenvalues were much larger for the TBRE than for the GOE. Cota {\it
et al.}~\cite{cot74} fitted the nearest--neighbour spacing
distribution to a Brody distribution, $P(s,\omega) = \alpha(\omega+1)
s^\omega \exp[-\alpha s^{\omega+1}]$ with $\alpha=\left\{\Gamma[(
\omega+2)/(\omega+1)]\right\}^{\omega+1}$, obtaining for the Brody
parameter the value $\omega \approx 0.80 \pm 0.05$. More recent
results by Flores {\it et al}~\cite{flo01} show that the semi--Poisson
distribution, $P(s) = 4 s \exp[-2s]$, gives a better fit than the
Brody distribution, if the levels of each member of the ensemble are
normalized first according to an individual Gaussian distribution
(spectral unfolding).

Numerical simulations using matrices of dimension $\sim 3000$ have
repeatedly found agreement between the fluctuation properties of the
EGE($k$) and the TBRE at the center of the spectrum on the one hand,
and the predictions of canonical RMT on the other. Therefore, it is
often taken for granted that a random $k=2$ part in the Hamiltonian is
sufficient to induce level fluctuations of canonical RMT type. This
point of view is at odds with the fact that many--body systems with
random few--body interactions (both for Fermions and Bosons) may
possess a high degree of order in the low--lying part of the
spectrum~\cite{joh98,bij99,joh99,bij00}. For instance, for the
TBRE with all the angular momentum couplings taken into
account explicitly, it was found that there is a statistical
preference for $J^\pi=0^+$ ground states despite the fact that these
states account only for a small fraction of Hilbert
space~\cite{joh98}. Moreover, these $J^\pi=0^+$ ground states are
separated by a gap from the lowest excited state. Other properties
like localization in Fock space~\cite{kap00} and odd--even binding
effects~\cite{pap02} show that pairing effects are robust properties
of two--body interactions. Such properties are usually understood in
terms of Hamiltonians which involve some collective behaviour. All
these results, based on Hamiltonians with random interactions,
contradict the basic philosophy of and the predictions based on
canonical RMT and suggest that the embedded ensembles may not always
yield results which coincide with canonical RMT.

\subsection{Bosonic Systems}
\label{sumb}

Early analytical results by Kota and Potbhare~\cite{kot80} indicated a
Gaussian spectral density both in the dilute and in the dense
limits. The dense limit which exists only for Bosons, is defined as $m
\to \infty$ for fixed $l$ and $k$. Manfredi~\cite{man84}, using
matrices of dimension $364$ ($l=4$, $m=11$), compared for the first
time the spectral fluctuation properties at the center of the spectrum
of EGE($k$) for Bosons with the results of canonical RMT. Using
spectral unfolding he concluded that there is no significant
difference between the two ensembles. Patel {\it et al}~\cite{pat00}
considered this problem in the dense limit, studying the case $l=5$
and $m=10$.  These authors constructed the ensemble--averaged
staircase function using a sixth order Gram--Charlier
expansion~(\ref{num:eq1}). They found excellent agreement with
canonical RMT results for the nearest--neighbour spacing distribution
(with a Brody parameter $\omega=0.95$) and for the
$\Delta_3$--statistic. These results led them to conclude that the
embedded ensembles possess generically (for Fermions and Bosons) the
spectral fluctuation properties of canonical RMT.

Recently it was found analytically, however, that the embedded
ensembles for Bosons are not ergodic in the dense
limit~\cite{asa01,asa02}. More precisely, the fluctuations of the
centroids and widths of the spectrum (in units of the average width)
do not vanish in this limit, but attain constant values
(cf. Subsection~\ref{low}). This fact implies that unfolding the
spectra by ensemble averaging or by spectral averaging will yield
different results. The Hilbert--space dimension for Bosons is given by
equation~(\ref{7}). It is clear that the effect can be best displayed
numerically for $l=2$.

\begin{figure}[t]
\centerline{\includegraphics[width=4.5cm,angle=90]{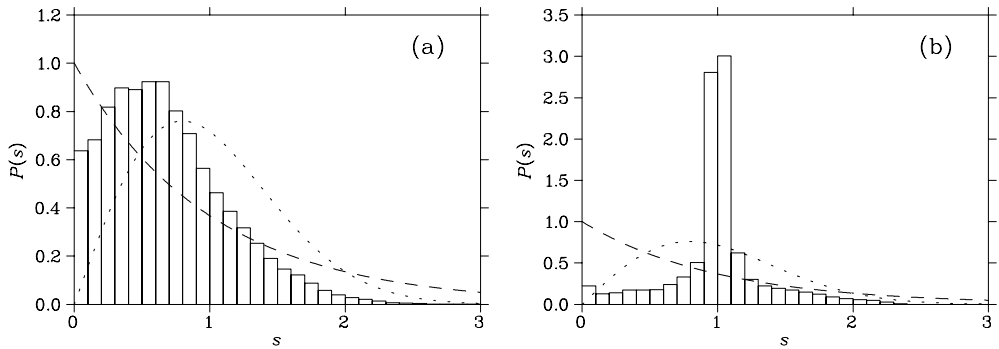}}
\caption{ Nearest--neighbour spacing distribution $P(s)$ obtained by
(a)~ensemble unfolding and (b)~spectral unfolding. The dotted curves
correspond to the Wigner surmise and the dashed ones to the Poisson
distribution. We note the different vertical scales used in the
frames. Taken from reference~\cite{asa02}.}
\label{f_1}
\end{figure}

\begin{figure}[t]
\centerline{\includegraphics[width=4.5cm,angle=90]{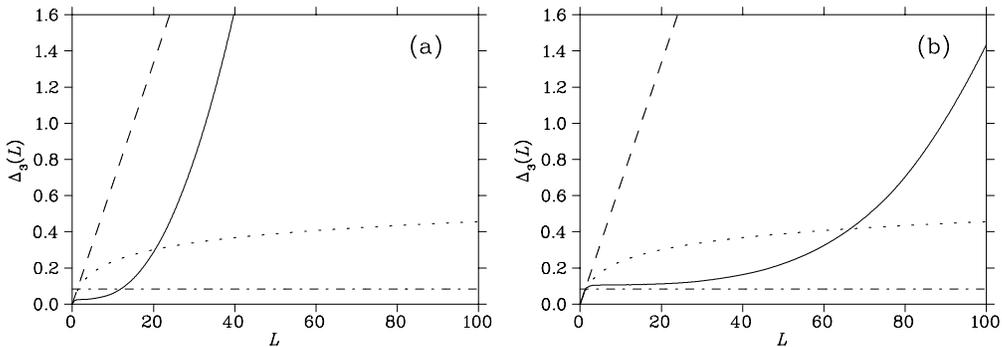}}
\caption{ $\Delta_3$--statistics (solid lines) measured at the center
of the spectrum after (a)~ensemble unfolding and (b)~spectral
unfolding. The dotted curve corresponds to the GOE results, the dashed
one to a Poissonian spectrum and the dotted--dashed one to a
picket--fence spectrum. Taken from reference~\cite{asa02}.}
\label{f_2}
\end{figure}

Figures~\ref{f_1} and~\ref{f_2} illustrate the non--ergodic behaviour
for $m=3000$ Bosons and the case of a two--body interaction (with
$1512$ members forming the ensemble) for the nearest--neighbour
spacing distribution and for the $\Delta_3$--statistic, both taken in
the centre of the spectrum. Spectral unfolding (ensemble unfolding)
was done by fitting a polynomial of degree $11$ to the staircase
function of each member of the ensemble (to the ensemble--averaged
staircase function, respectively). For ensemble unfolding, the
individual spectra were not recentered or rescaled since there is now
no theoretical support for this procedure. In fact, rescaling and
recentering the spectra would yield a non--Gaussian average level
density. Interestingly, in the case of spectral unfolding the
nearest--neighbour spacing distribution is dominated by a large peak
centred at $s = 1$. This suggests that individual spectra have an
almost constant level spacing, i.e., are close to spectra of
picket--fence type. This is further supported by the large plateau
observed in the plot for the $\Delta_3$--statistic.

\begin{figure}
\centerline{\includegraphics[width=11.2cm]{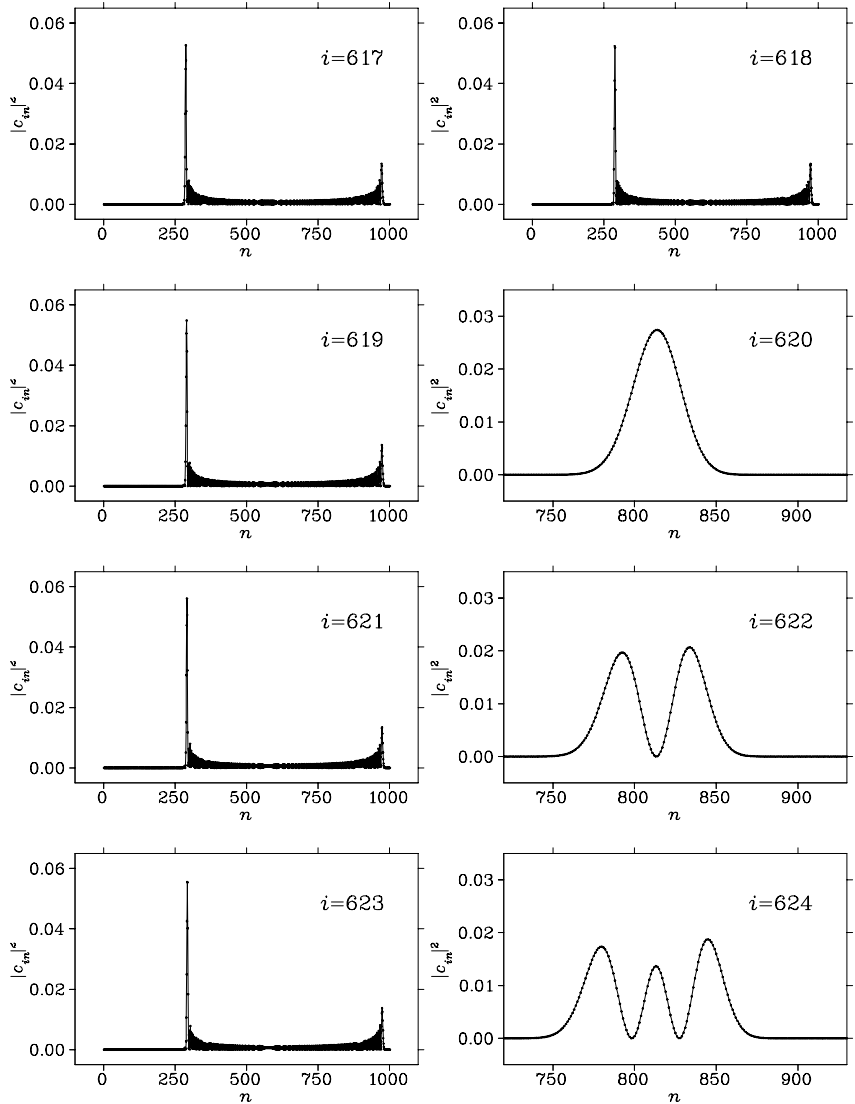}}
\caption{ Intensities $|c_{in}|^2$ of a sequence of eigenvectors in
the many--body basis in the vicinity of a kink in the staircase
function. The two overlapping segments of nearly equidistant levels
are easily distinguished by the structure of the corresponding
eigenfunctions. Taken from reference~\cite{asa02}.}
\label{f_3}
\end{figure}

For the same case ($l = 2, m = 3000$) the structure of the
eigenfunctions was also investigated~\cite{asa01,asa02}. It was found
that individual spectra consist generically of a superposition of
independent sequences of levels with constant spacings as if there was
some symmetry present in the system. When a new sequence appears, the
old and the new sequence are interlaced, and the level spacing is no
longer constant for a while. In the staircase function, this is seen
as a sudden change in the average slope, i.e., as a kink. The first
few eigenstates of the new sequence are localized in Hilbert space,
even though they may correspond to rather highly excited states of the
spectrum. This is illustrated in figure~\ref{f_3}, where the squares
of the coefficients $c_{in}$ of the eigenvector expansion in an
ordered many--body basis $|\mu_n\rangle$ are plotted
(cf. reference~\cite{asa02}).

These numerical results show that in the dense limit, the Bosonic
EGE($k$) differ significantly from the canonical ensembles of RMT. In
the case $l=2$ it has been established that the difference is due to
the exact integrability of the system in the semiclassical
limit~\cite{ben02}. Essentially, this case corresponds to the
quantization of a classical two--degrees of freedom system which
possesses two independent integrals of motion, the energy and the
number of particles. For arbitrary values of $l$, the corresponding
classical system has $l$ degrees of freedom. As $l$ increases, chaotic
trajectories appear in phase space, typically for energies around the
center of the spectrum, and there is a corresponding transition in the
spectral fluctuations of the quantum system.

In concluding this Section we mention that the spectral properties at
the edge of the spectrum have also been investigated for Bosons. Using
a random interaction in the framework of the Interacting Boson Model,
Bijker and Frank have demonstrated the emergence of collective motion
in the low--lying part of the spectrum, and have studied the
dependence of collective motion on the rank $k$ of the
interaction~\cite{bij99,bij00,bij00a}. When the number of interacting
Bosons $m$ is large enough with respect to $k$, they find a
preponderance of $J^\pi=0^+$ ground states as well as definite
evidence for the appearance of vibrational and rotational
structures. For $m \sim k$ they do not obtain an indication of
vibrational or rotational band structures.

\section{The Method of Replicated Variables}
\label{rep}

Verbaarschot and Zirnbauer~\cite{ver84} used the replica trick
developed in statistical mechanics for the study of spin glasses and
of Anderson localization to investigate spectral properties of the GOE
and of EGOE($k$). Using the replica trick, one writes the observable
under study as the logarithmic derivative of a suitably chosen
generating function $Z$, replaces $\log Z$ by $\lim_{n \rightarrow 0}
[ Z^n - 1 ]$, evaluates $\overline{Z^n}$ for positive integer values
of $n$ and takes the limit $n \rightarrow 0$ of the resulting
expression. Similar to what happens in the supersymmetry approach
described in Section~\ref{eig} below, averaging $Z^n$ yields in the
exponent of $Z^n$ a quartic term in the integration variables which
contains the matrix $A^{(k)}_{\mu \nu; \rho \sigma}(\beta)$ defined in
equation~(\ref{10}). The authors introduce matrix elements of the
operators $\psi^{\dagger}_{k, \alpha}$ which connect the $m$--particle
states with the $(m - k)$--particle states and perform the
Hubbard--Stratonovich transformation in this mixed representation.

They use the saddle--point approximation for the one--point function
and find a semicircle for the spectral density. The moments method
having shown that the semicircle does not apply for all values of
$(m,k)$, this points to the need to study the loop expansion. It is
shown that there are terms in the loop expansion which correct the
semicircle. However, it is not possible to evaluate all such
corrections.

The saddle--point approximation is then applied to the two--point
function within the same mixed representation as used for the
one--point function. The saddle--point solution allows for the
existence of a Goldstone mode. This mode carries the variable $s =
(E_1 - E_2)/d$ where $d$ is the mean level spacing. It is with respect
to this variable that fluctuation properties of the spectrum are
measured. The two--point function is evaluated using an expansion in
inverse powers of $s$, the point being that the stiffness of the GOE
spectrum relates to the occurrence of the factor $s^{-2}$ in the
average two--point function. Difficulties and, in fact, uncertainties
arise from the need to integrate over the massive (i.e., the
non--Goldstone) modes. With this proviso, it is found that the leading
terms in the loop expansion combine to yield the value $- (17/2) (1 /
(\pi^2 s^2))$ for the density--density correlation function. This
result has to be compared with the GOE value $- (1 / (\pi^2
s^2))$. Led by early numerical results reviewed in Section~\ref{num},
the authors believe that they should have obtained the GOE result and
speculate that the difference is due to high--order terms in the loop
expansion. They show that if the density--density correlation function
is $\propto s^{-2}$ then stiffness of the spectrum is implied.

Results obtained more recently~\cite{ben01a,ben01b,asa01,asa02} and
reviewed in Sections~\ref{eig} and~\ref{bin} cast some doubt on the
assertion that the spectral fluctuation properties of EGE($k$)
coincide, for all values of $k$, with those of the canonical
ensembles, at least in the limit $l \rightarrow \infty$. This would
imply that higher--order loop corrections in the approach of
reference~\cite{ver84} could modify the factor $(17/2)$ in a
$k$--dependent fashion.

\section{The Matrix of Second Moments}
\label{eig}

A novel approach to determine spectral properties of EGE($k$) was
developed in references~\cite{ben01a,ben01b,asa01,asa02}. We describe
this approach here for the case of Fermions. The case of Bosons can be
treated in complete analogy. Starting point is the observation that
the matrix elements $\langle \mu | V_k | \nu \rangle$ of the
stochastic operator $V_k$ are themselves also random variables and
have a Gaussian probability distribution with mean value zero.
Therefore, all properties of the embedded ensembles are completely
determined by the matrix $A^{(k)}_{\mu \nu; \rho \sigma}(\beta)$ of
second moments defined by
\be 
A^{(k)}_{\mu \nu; \rho \sigma}(\beta) = \overline{\langle \mu |
V_k(\beta) | \sigma \rangle \langle \rho | V_k(\beta) | \nu \rangle}
\ .
\label{10}
\ee
The idea is to extract information on EGE($k$) by studying the
properties of the matrix $A^{(k)}_{\mu \nu; \rho \sigma}(\beta)$. As 
in the case of canonical random matrices, we use the labels $\beta=1$ 
and $\beta=2$ for the orthogonal and unitary ensemble, 
respectively. Performing the average yields
\ba
A^{(k)}_{\mu \nu; \rho \sigma}(\beta) &=& \sum_{\alpha \delta} \biggl [
\langle \mu | \psi^{\dagger}_{\alpha} \psi_{\delta} | \sigma \rangle
[ \langle \rho | \psi^{\dagger}_{\delta} \psi_{\alpha} | \nu \rangle
+ \delta_{\beta 1} \langle \rho | \psi^{\dagger}_{\alpha}
\psi_{\delta} | \nu \rangle ] \biggr ] \nonumber \\
&=& \sum_{\alpha \delta} \biggl [ \langle \mu | \psi^{\dagger}_{\alpha}
\psi_{\delta} | \sigma \rangle [ \langle \rho | \psi^{\dagger}_{\delta}
\psi_{\alpha} | \nu \rangle + \delta_{\beta 1} \langle \nu |
\psi^{\dagger}_{\delta} \psi_{\alpha} | \rho \rangle ] \biggr ] \ .
\label{11}
\ea
In the last of equations~(\ref{11}) we have used the reality of the
matrix element $\langle \nu | \psi^{\dagger}_{\delta} \psi_{\alpha} |
\rho \rangle$.

\subsection{Duality. Eigenvector Expansion}
\label{dual}

In the unitary case, there is a connection between the matrices
$A^{(k)}(2)$ for the $k$--body interaction and $A^{(m-k)}(2)$ for the
$(m-k)$--body interaction. The relation is referred to as duality and
reads
\be
A^{(k)}_{\mu \nu; \rho \sigma}(2) = A^{(m-k)}_{\mu \sigma; \rho
\nu}(2) \ .
\label{12}
\ee
We note the difference in the sequence of indices $\{ \mu \nu \rho
\sigma \}$ on the two sides of this equation. The proof of
equation~(\ref{12}) rests on the fact that for every $m$--Fermion
state $| \mu \rangle$ and for every operator $\psi_{k, \alpha}$ with
$\psi_{k, \alpha} | \mu \rangle \neq 0$, there exists a uniquely
defined operator $\psi_{(m - k), \gamma}$ such that $\psi_{(m - k),
\gamma} \psi_{k, \alpha} | \mu \rangle = | 0 \rangle$, the vacuum
state. We emphasize that the duality relation has nothing to do with
particle--hole symmetry. It applies likewise to Bosons and can be
extended to the case $\beta = 1$. The duality relation obviously
assigns a special role to the case $2k = m$. This is the reason why
spectral properties of EGE($k$) change at $2k = m$.

For $2m > l$, the available single--particle states are more than half
filled, and it is tempting to use particle--hole symmetry to simplify
the algebra. Rewriting $V_k$ in terms of hole operators and bringing
the hole creation operators up front, one creates a sum of $k'$--body
interactions with $k' = 1,2,\ldots,k$. This is why particle--hole
symmetry is of limited use only, except for the dilute limit.

A very useful tool in the analysis of spectral properties of EGE($k$)
is the eigenvector expansion of the matrix $A^{(k)}$. We take the
unitary case, $\beta = 2$, for Fermions and consider the matrix
$A^{(k)}$ of second moments as a matrix in the product space $\{\mu
\nu\}$. In this space, $A^{(k)}(2)$ is a Hermitean matrix and can,
therefore, be diagonalized. The eigenvalue equation reads
\be
\sum_{\rho \sigma} A^{(k)}_{\mu \nu; \rho \sigma}(2) C^{(s a)}_{\sigma
\rho} = \Lambda^{(s)}(k) C^{(s a)}_{\mu \nu} \ .
\label{13}
\ee
The index $s = 0,1,\ldots$ labels different eigenvalues
$\Lambda^{(s)}$ and the index $a$ labels degenerate eigenvectors. The
eigenvectors are orthogonal. We choose the normalization
\be
\sum_{\mu \nu} C^{(s a)}_{\mu \nu} C^{(t b)}_{\nu \mu} = N \delta_{s
t} \delta_{a b} \ .
\label{14}
\ee
Provided that the eigenvectors form a complete set, the matrix
$A^{(k)}$ can be expanded in the form
\be
A^{(k)}_{\mu \nu; \rho \sigma}(2) = \frac{1}{N} \sum_{s=0}^{m-k}
\sum_a \Lambda^{(s)}(k) C^{(s a)}_{\mu \nu} C^{(s a)}_{\rho \sigma}
\ .
\label{15}
\ee
In writing equation~(\ref{15}) we have anticipated the fact that for 
$s > (m-k)$, the eigenvalues $\Lambda^{(s)}$ vanish. Knowledge of the
expansion~(\ref{15}), i.e., of the eigenvectors and eigenvalues of
$A^{(k)}$, makes it possible to calculate the low moments of $V_k$.
Moreover, using the expansion~(\ref{15}) in the framework of
supersymmetry allows for the use of the Hubbard--Stratonovich
transformation. The orthogonal case is treated analogously. Combining
the results for $\beta = 1,2$, we obtain
\be
A^{(k)}_{\mu \nu; \rho \sigma}(\beta) = \frac{1}{N} \sum_{s=0}^{m-k}
\sum_a \Lambda^{(s)}(k) [ C^{(s a)}_{\mu \nu} C^{(s a)}_{\rho \sigma}
+ \delta_{\beta 1} C^{(s a)}_{\mu \rho} C^{(s a)}_{\nu \sigma}] \ .
\label{15a}
\ee

To appreciate the significance of the eigenvector expansion
equation~(\ref{15}), it is useful to recall the form of the second
moment of the GUE Hamiltonian $H_{\mu \nu}$,
\be
\overline{H_{\mu \sigma} H_{\rho \nu}} = \frac{\lambda^2}{N}
\delta_{\mu \nu} \delta_{\rho \sigma} \ . 
\label{16}
\ee
We have put $v^2 = 1$ for EGE($k$) and, by analogy, put $\lambda =
1$ for the GUE. The fact that GUE coincides with EGUE($m$) then
implies that for $k = m$, equation~(\ref{15}) reduces to
equation~(\ref{16}).  Put differently, the one non--vanishing
eigenvalue of the GUE is $\Lambda^{(0)}(m)$, and the associated
non--degenerate eigenvector is $C^{(0)}_{\mu \nu} = \delta_{\mu
\nu}$. All other orthogonal eigenvectors belong to eigenvalue
zero. The comparison shows that the eigenvalue expansion
equation~(\ref{15}) is the natural generalization to EGUE($k$) of
equation~(\ref{16}) for the second moment of the GUE Hamiltonian.
The Kronecker delta's in equation~(\ref{16}) express the unitary
invariance of GUE. The fact that for $k < m$ the sum over $s$ in
equation~(\ref{15}) extends up to $(m-k)$, with eigenvectors $C^{(s
a)}_{\mu \nu}$ which differ from Kronecker delta's, is due to the fact
that $V_k$ does not possess this unitary invariance for $k < m$. We
remark in parenthesis that the eigenvalue in equation~(\ref{16})
obviously has value unity, while it turns out that for $m = k$, the
eigenvalue $\Lambda^{(0)}(m)$ in equation~(\ref{15}) has the value $l
\choose k$. The difference is due to the fact that when we use second
quantization to write GUE = $V_m$ and work out $A^{(m)}(2)$, we have
to count the number of holes which is $l \choose k$. In the appendix
of reference~\cite{ver84}, results closely related to
equations~(\ref{10}) to (\ref{15a}) were derived.

It remains to determine the form of the eigenvectors and eigenvalues.
The eigenvectors $C^{(s a)}_{\mu \nu}$ have the form
\be
C^{(s a)}_{\mu \nu} = \langle \mu | \psi^{\dagger}_{s, \alpha}
\psi_{s, \gamma} | \nu \rangle \ ,
\label{17}
\ee
where $a$ stands for the set $(\alpha, \gamma)$. The form
equation~(\ref{17}) applies for all $s = 0,1,\ldots,m$ independently of
the actual value of the eigenvalue $\Lambda^{(s)}(k)$. Whenever two
or more single--particle indices in the sets $\alpha$ and $\gamma$
coincide, special attention is required~\cite{ben01b}. For this
reason, the number $D^{(s)}$ of orthonormal degenerate eigenvectors
$C^{(s a)}_{\mu \nu}$ belonging to fixed $s$ (and, thus, the
dimension of the corresponding subspace of Hilbert space) are given
by
\be
D^{(s)} = {l \choose s}^2 - {l \choose s-1}^2 \ .
\label{18}
\ee 
It follows that $\sum_{s=0}^m D^{(s)} = {l \choose m}^2$, the
dimension of the product space $\{\mu \nu\}$, so that the eigenvectors
form a complete set. The eigenvalues are given by
\be 
\Lambda^{(s)}(k) = {m-s \choose k} {l-m+k-s \choose k} \ .
\label{19}
\ee
We observe that $\Lambda^{(s)}(k) = 0$ for $s > (m-k)$. This fact
limits the sums over $s$ in equations~(\ref{15},\ref{15a}). Using the
eigenvector expansion in the duality relation, equation~(\ref{12}),
generates useful identities.

For Bosons, the eigenvalue expansion equation~(\ref{15a}) applies likewise.
The eigenvectors have the form given in equation~(\ref{17}), with $\psi$
replaced by $\chi$, and are normalized as in equation~(\ref{14}), with $N$
replaced by $N_B$. The dimension $D^{(s)}_B$ of the subspace spanned
by degenerate eigenvectors characterized by $s$ with $s = 0,1,\ldots,
m$ is given by $D^{(0)}_B =1$ and, for $s \geq 1$, by
\be
D^{(s)}_B = {l+s-1 \choose s}^2 - {l+s-2 \choose s-1}^2 \ .
\label{18a}
\ee
Again, we have $\sum_{s = 0}^m D^{(s)}_B = N_B^2$, showing that the
eigenvectors form a complete set. The eigenvalues are given by
\be
\Lambda^{(s)}_B(k) = {m-s \choose k} {l+m+s-1 \choose k} \ .
\label{19a}
\ee
The eigenvalues vanish for $s > m - k$, in keeping with the GUE result
where for $k = m$ the only non--vanishing eigenvalue belongs to $s =
0$.

\subsection{Group--Theoretical Aspects}
\label{grou}

Duality and the eigenvector decomposition equation~(\ref{15a}) apply to
both Fermions and Bosons. A group--theoretical analysis~\cite{plu02}
shows that both these results and much of the structure displayed in
Subsection~\ref{dual} apply much more widely and are rooted in
symmetry properties of the embedded ensembles. The analysis also
gives a deeper meaning to the concept ``embedded ensemble''. For
simplicity, we confine ourselves to the unitary case although all
results apply likewise to the orthogonal one. We denote by
$\alpha^{\dagger}_j$ the creation operator for a Fermion or a Boson
in single--particle state $j$ and by $\Psi^{\dagger}_{k \alpha}$
the creation operator of a normalized state containing $k$ particles
(Fermions or Bosons). This normalized state need not be a Slater
determinant or a product state (as is the case for the operators
$\psi_{k \alpha}$ and $\chi_{k \alpha}$ defined in Section~\ref{defi})
but may be a linear combination of such states.

Three symmetry groups are relevant for the embedded ensembles. These
are (i) the group SU($l$) of unitary transformations of the $l$
degenerate single--particle states $| j \rangle$ with $j = 1, \ldots,
l$; (ii) the group U($N_k$) of unitary transformations of the
$k$--body interaction where $N_k$ is the dimension of the Hilbert
space containing $k$ particles (Fermions or Bosons); (iii) the group
U($N_m$) of unitary transformations of the Hilbert space containing
$m$ particles, with $N_m$ the dimension of this space. As we shall
see, the group SU($l$) governs the embedding, the group U($N_k$) is
obviously the symmetry group of EGUE($k$), and U($N_m$) is the
symmetry group of the GUE.

An element $u \in$ SU($l$) generates a unitary transformation of the
single--particle states and a corresponding transformation $T(u)
\alpha^{\dagger}_j T^{\dagger}(u)$ of the creation operators
$\alpha^{\dagger}_j$. Under $u$, the operators $\Psi^{\dagger}_{t \alpha}$
transform according to the irreducible representation $D^{f_t}_{\alpha
\gamma}(u)$,
\be
T(u) \Psi^{\dagger}_{t \alpha} T^{\dagger}(u) = \sum_{\gamma}
D^{f_t}_{\alpha \gamma}(u) \Psi^{\dagger}_{t \gamma} \ . 
\label{20a}
\ee
Here $t$ is an integer which may take the values $m, k$, or $(m - k)$.
For Fermions (Bosons), the matrices $D$ are in essence totally
antisymmetrized (symmetrized, respectively) powers of $u$. 

In analogy with the familiar fractional--parentage technique, we can
expand the $m$--particle states $| m \alpha \rangle = \Psi^{\dagger}_{m
\alpha} | 0 \rangle$ into products of states containing $k$ and $(m -
k)$ particles, respectively,
\be
| m \alpha \rangle = {m \choose k}^{-1/2} \sum_{\gamma \delta}
\Psi^{\dagger}_{k \gamma} | (m-k) \delta \rangle C^{f_m \alpha}_{f_k
\gamma f_{(m-k)} \delta} \ .
\label{20b}
\ee
The coefficients $C^{f_m \alpha}_{f_k \gamma f_{(m-k)} \delta}$ are
the coefficients of fractional parentage or, equivalently, the
Clebsch--Gordan coefficients for the coupling $[(f_k f_{(m-k)}) f_m]$
of the irreducible representations $f_k$ and $f_{(m-k)}$ to $f_m$.

The product $\Psi^{\dagger}_{k \gamma} \Psi_{k \delta}$ of operators
which appears in the definition of the $k$--body embedded ensemble
transforms according to the direct product of irreducible
representations $D^{f_k}(u)$ and $D^{\overline{f}_k}(u)$. The product
can be reduced to a direct sum of irreducible representations
$D^{g_b}(u)$. This defines a set of basic $k$--particle interactions
$B_k(b \alpha)$ which are Hermitean, transform according to the
irreducible representation $D^{g_b}(u)$, and are given by
vector--coupling $\Psi^{\dagger}_{k \gamma}$ and $\Psi_{k \delta}$
via the Clebsch--Gordan coefficient $C^{g_b \alpha}_{f_k \gamma
\overline{f}_k \delta}$. Using the Wigner--Eckart theorem, we can
write the matrix element $\langle m \gamma | B_k(b \alpha) | m
\delta \rangle$ as the product of a reduced matrix element $\langle m
|| B_k(b) || m \rangle$ and the Clebsch--Gordan coefficient $C^{g_b
\alpha}_{f_m \gamma \overline{f}_m \delta}$.

The interaction $V_k$ can now be rewritten as a sum over the operators
$B_k(b \alpha)$. The coefficients are uncorrelated Gaussian
distributed random variables. Using this form and the Wigner--Eckart
theorem, one finds for the second moment of the matrix elements
$\langle m \gamma | V_k | m \delta \rangle$ an expression which is
equivalent to the eigenvalue decomposition equation~(\ref{15a}). The role
of the eigenvectors is played now by the Clebsch--Gordan coefficients,
and the role of the eigenvalues is taken by the squares of the reduced
matrix elements $\langle m || B_k(b) || m \rangle$. The index $b$
coincides with the summation index $s$ in equation~(\ref{15a}). The duality
relation also follows from the Wigner--Eckart theorem. Thus, all the
relations derived explicitly for Fermions and Bosons are seen to
follow from group--theoretical considerations.

These insights allow for a generalization of the concept of an embedded
ensemble which brings the underlying concepts to the fore most clearly.
We consider an arbitrary compact simple Lie group $G$ and two independent
systems labelled $k$ and $(m-k)$ whose basic states $| f_k \alpha
\rangle$ and $| f_{(m-k)} \gamma \rangle$ transform according to the
irreducible representations $D^{f_k}(g)$ and $D^{f_{(m-k)}}(g)$ of the
group $G$, with $g \in G$. Now let us assume that a non--trivial
interaction $V_k$ of GUE type occurs only in system $k$. The embedding
of this interaction into a space of different dimension is accomplished
by projecting the product states $| f_k \alpha \rangle | f_{(m-k)}
\gamma \rangle$ onto the subspace of states which transform according to
the irreducible representation $D^{f_m}(g)$ which is contained in the
direct product $D^{f_k}(g) \times D^{f_{(m-k)}}(g)$. Let the associated
projection operator be denoted by $I(f_m)$. The embedded ensemble is
then defined by the Hamiltonian $I(f_m) V_k I(f_m)$. This definition
comprises the essence of the group--theoretical extension of the idea
of an embedded ensemble. It is independent of the existence of Fermions
and Bosons and relies only on group--theoretical concepts.

Group--theoretical arguments can also be used to isolate that part of
$V_k$ which is invariant under $U(N_m)$, and to investigate symmetries
of the generating functional in the supersymmetry approach~\cite{plu02}.
The part of $V_k$ which is invariant under $U(N_m)$ transforms either
like the GUE or like a multiple of the unit operator. Both possibilities
actually occur in the decomposition. The latter, taken by itself, would
cause Poissonian level statistics. It carries particularly large weight
for $k \ll m$.

\subsection{Moments of $V_k$}
\label{low}

The eigenvector expansion equation~(\ref{15a}) and the orthogonality
relations equation~(\ref{14}) make it possible to calculate low moments
and variances of $V_k$ without resorting to the dilute limit $l \gg
m$, at least for $\beta = 2$. Three observables of interest are
  \ba
S &=& \frac{\overline{((1/N) {\rm tr} V_k(\beta))^2}}{(1/N) {\rm tr}
\overline{(V_k(\beta))^2}} \ , \nonumber \\
\nonumber \\
R &=& \frac{\overline{((1/N) {\rm tr} (V_k(\beta))^2)^2} \ - \ ( \
(1/N) {\rm tr} \overline{(V_k(\beta))^2} \ )^2}{( \ (1/N) {\rm tr}
\overline{ V_k(\beta))^2} \ )^2} \ , \nonumber \\
\nonumber \\
\kappa &=& 2 + Q = \frac{(1/N) {\rm tr} \overline{(V_k(\beta))^4}}{(
\ (1/N) {\rm tr} \overline{V_k(\beta))^2} \ )^2} \ .
\label{20}
\ea The ratio $S$ measures the fluctuations of the center of the
spectrum in units of the average width of the spectrum. The ratio $R$
measures the relative fluctuation of the width of the spectrum. The
parameter $\kappa$ is the kurtosis, and $Q$ marks the difference in
spectral shape between the semicircle ($Q = 0$) and the Gaussian ($Q =
1$). Explicit values for $S, R, Q$ are given in
references~\cite{ben01b,asa02}, both for Fermions and for Bosons. The
result for $R$ was first obtained by French~\cite{fre73}. Both $S$ and
$R$ vanish in the limit $l \to \infty$. However, for fixed values for
$k$ and $m/l$, both ratios decrease very slowly (with inverse powers
of the logarithm) with increasing dimension $N$ of the matrices. This
fact is at the root of the difficulty in obtaining reliable spectral
information on EGE($k$) from numerical simulations.

For the shape of the spectrum, the quantity $Q$ is of primary interest.
It is explicitly given~\cite{ben01b} in terms of the eigenvalues
$\Lambda^{(s)}(k)$ and the dimensionalities $D^{(s)}$ introduced above.
In the limit $l \rightarrow \infty$, $Q$ vanishes if $2k > m$ with both
$k$ and $m$ fixed, and with both $k/m$ and $m/l$ fixed, while $Q$
attains a finite and non--zero value for $2k \leq m$ with both $k$ and
$m$ fixed, and with both $k$ and $m/l$ fixed. This suggests that the
transition from the semicircular to Gaussian shape takes place for $2k
\leq m$. (It would take the study of higher moments to make this
conclusion unambiguous). As mentioned above, the critical role played
by the value $2k = m$ is attributed to duality.

For Bosons in the dense limit, the ratios $S$ and $R$ do not vanish as
$m \to \infty$. Thus, the fluctuations of the centroids and widths of
individual spectra do not vanish asymptotically. Hence, the Bosonic
ensembles are not ergodic in the dense limit $m \to \infty$ with $k$
and $l$ fixed.

\subsection{Supersymmetry Approach}
\label{sup}

The eigenvector expansion equation~(\ref{15a}) makes it possible to apply
the supersymmetry approach. To see this, we recall that when one uses
supersymmetry to calculate properties of the GUE, equation~(\ref{16}) is
extremely helpful. Indeed, after averaging the generating function over
the ensemble, there appears in the exponent a term which is quartic in
the commuting and anticommuting integration variables. Owing to
equation~(\ref{16}), this quartic term can be written as the square of
a bilinear form, and the latter is handled with the help of the
Hubbard--Stratonovich (HS) transformation. Because of equation~(\ref{15a}),
a similar situation arises for EGE($k$). The form of this equation
implies that the quartic term in the integration variables obtained
after ensemble--averaging, is a sum of squares of bilinear forms, and
the HS transformation again can be used. There is a price, however:
Whereas only a single graded matrix $\sigma$ is needed to carry out
the HS transformation for the GUE, the EGE($k$) requires the
introduction of as many $\sigma$--fields as there are independent
eigenvectors to non--zero eigenvalues in equation~(\ref{15a}). This
number rises steeply as $m$ increases from $k = m$, see
equation~(\ref{18}).

The saddle--point approximation to the integration over the
$\sigma$--fields yields for EGE($k$) a semicircle for the spectral
density, and universal Wigner--Dyson spectral fluctuations, just like
for the GUE. The loop expansion, generated by expanding the
$\sigma$--fields around the saddle--point solution, serves as a test
of the saddle--point approximation. For the GUE, all terms in the loop
expansion vanish asymptotically for $N \rightarrow \infty$. For
EGE($k$), it is not possible technically to go beyond the lowest
non--vanishing term of the loop expansion. For the spectral density,
one finds that for $2k > m$, this term vanishes asymptotically for
$l \rightarrow \infty$. This fact reinforces the conclusion in
subsection~\ref{low} that the spectral density has semicircular
shape for $2k > m$. For $2k \leq m$, finite corrections appear the
form of which is consistent with the value of $Q$ found in
subsection~\ref{low}. For the spectral fluctuations (the two--point
function), the situation is more ambiguous. The lowest--order loop
correction vanishes asymptotically for all values of $k$ albeit it
does so ever more slowly as $k$ decreases from $k = m$. This might
be consistent with universal Wigner--Dyson level statistics for all
values of $k$. We note, however, that the lowest--order loop
correction also produces non--universal terms. Such terms might also
arise in higher order and not vanish asymptotically.

In summary, the supersymmetry approach yields useful information
especially for $2k > m$. It lends independent support to the
conclusion that the change from semicircular to Gaussian spectral
shape sets in as $k$ decreases and passes through the value $2k = m$.
For the spectral fluctuations, it is consistent with but not
necessarily in support of Wigner--Dyson statistics for all values of
$k$. However, for the physically most interesting case $k = 2$, it
yields very little information.

\section{Binary Correlation Approximation}
\label{bin}

None of the analytical approaches described above has been able to
yield definitive information on the spectral fluctuation properties of
EGE($k$) for the physically interesting case $k \ll m$. There exists,
however, another approach first introduced by Mon and French for the
calculation of the one--point function~\cite{mon75}. This is the
binary correlation approximation which applies in the dilute limit $k
\ll m \ll l$, $m/l \to 0$ as $m,l \to \infty$~\cite{bro81}. The binary
correlation approximation can be generalized to investigate the
two--point function~\cite{ben01a,ben01b}. We first review the
calculation of the one--point function of the EGUE($k$) in the version
of Verbaarschot and Zirnbauer~\cite{ver84}, and then present that of
the two--point function~\cite{ben01a,ben01b}. We note that in the
dilute limit the distinction between Fermions and Bosons is physically
irrelevant, both cases yielding the same results. For definiteness, we
consider the case of Fermions. To be free of singularities, we choose
in this Section $v^2=[\Lambda^{(0)}(k)]^{-1}$ which implies a bounded
spectrum of unit width.

\subsection{One--Point Function}
\label{1point}

Following Verbaarschot and Zirnbauer~\cite{ver84}, we expand the
one--point function in a power series in $V_k$,
\begin{equation}
\overline{g(z)} \equiv {1 \over N} {\rm tr}
  \biggl( \overline{1 \over z - V_k} \biggr ) 
= {1 \over N} \sum_{p = 0}^{\infty} {1 \over z^{2p+1}} 
  {\rm tr} \biggl ( \overline{V_k^{2p}} \biggr ) \ ,
  \label{bin:eq2}
\end{equation}
where $z=E-i\eta$, $E$ is the energy and $\eta > 0$ is an
infinitesimal increment. In the second equation in~(\ref{bin:eq2}), we
have interchanged the summation and the average over the ensemble, and
we have used the Gaussian distribution of $V_k$ which implies that
only even powers of $V_k$ contribute. The ensemble average is carried
out using Wick contractions. Each contracted pair of $V_k$'s is
evaluated in the dilute limit, i.e., replaced by $v^2 \,
\Lambda^{(0)}(k) = 1$. In the term of order $p$, there are $(2p-1)!!$
different ways of pairwise contracting the $V_k$'s. The result is
\begin{equation}
\overline{g(z)} = \sum_{p = 0}^{\infty} 
  \frac{1}{z^{2p+1}} (2p-1)!! = \frac{1}{z} \sum_{p = 0}^{\infty} 
  \frac{1}{p!} \biggl[ \frac{1}{2z^2} \biggr]^p (2p)! \ .
\label{bin:eq3}
\end{equation}
This expression is evaluated using the technique of Borel resummation.
We use the identity $n ! = \int_0^\infty e^{-t} t^n dt $ for the
factor $(2p)!$. We interchange the summation and the integration. The
sum over $p$ yields an exponential function, thus leading to
\begin{equation}
\overline{g(z)} = \frac{1}{z} \int_0^\infty dt 
  \exp{ \biggl[ \frac{t^2}{2z^2} -t \biggr]} \ .
\label{bin:eq4}
\end{equation}
Equation~(\ref{bin:eq4}) converges provided ${\rm Re} [z^{-2}] <
0$. Writing $z=-i|z|e^{i\phi}$, we see that this condition is
fulfilled for $-\frac{\pi}{4} < \phi < \frac{\pi}{4}$. The integral is
evaluated as follows. First, we choose as contour the straight line
that joins the origin and $z$. Putting $t=\tau e^{i\phi}$, we rotate
this contour so that it comes to lie on the real axis. A new (real)
integration variable $u=\tau/|z|$ is introduced. The resulting
integral converges for all $z$, and is evaluated in a straightforward
manner, yielding~\cite{ver84}
\begin{equation}
\overline{g(z)} = i \sqrt{\pi\over2} \exp \biggl[-\frac{z^2}{2} \biggr] 
  \, {\rm erfc}\biggl[\frac{iz}{\sqrt{2}} \biggr] \ .
\label{bin:eq5}
\end{equation}
Taking the limit $\eta\to 0$ and recalling that the mean level density
is given by $\overline{\rho(E-i\eta)}=\pi^{-1}{\rm
Im}[\overline{g(E-i\eta)}]$ we obtain the Gaussian form for the average
level density
\begin{equation}
\overline{\rho(E)} = (2\pi)^{-1/2} \exp{\biggl[-\frac{E^2}{2}\biggr]}
  \ .
\label{bin:eq6}
\end{equation}

\subsection{Two--Point Function}
\label{2point}

Our presentation differs slightly from that in reference~\cite{ben01b}
and takes account of recent developments~\cite{sre02}. Again, we work
in the dilute limit and consider the quantity
\begin{equation}
R_2(z_1,z_2) = \frac{\overline{g(z_1) g(z_2)}} {\overline{g(z_1)}
  \cdot \overline{g(z_2)}} - 1 \ ,
\label{bin:eq7}
\end{equation}
where $z_1^\pm = E^\pm + \varepsilon/2$ and $z_2^\pm = E^\pm -
\varepsilon/2$, with $E$ and $\varepsilon$ the mean value and the
difference of the energy arguments of the two Green functions,
respectively. An upper plus (minus) sign denotes an infinitesimal
positive (negative) imaginary energy increment, respectively. We are
interested in values of $\varepsilon$ which are of the order of the
mean level spacing and, thus, small compared to unity (the width of
the spectrum). Thus, we have approximately $|z_1| \sim |z_2|$. In
order to obtain the connected part of the density--density correlator,
$z_1$ and $z_2$ must have imaginary energy increments of opposite
signs.

Proceeding as in the case of the one--point function, we expand
$R_2(z_1,z_2)$ in powers of $V_k$. Using equation~(\ref{bin:eq2}) we
have
\begin{equation}
\overline{g(z_1) g(z_2)} = \frac{1}{N^2} \sum_{p = 0}^{\infty} \sum_{q
  = 0}^{\infty}\frac{1}{z_1^{p+1} z_2^{q+1}}\overline{{\rm tr}
  [(V_k)^p] \ {\rm tr} [(V_k)^{q}]} \ .
\label{bin:eq8}
\end{equation}
The Gaussian distribution of the $V_k$ implies that $p+q$ must be
even.  Using Wick contraction we obtain two types of
contributions. The members of the contracted pair either occur under
the same trace, or involve both traces. We refer to the latter as
cross--contractions. Reordering terms, we obtain
\begin{equation}
\fl
\overline{g(z_1) g(z_2)} = \frac{1}{N^2} \sum_{s = 0}^{\infty} \,
  \sum_{p,q = 0}^{\infty} \, \frac{1}{z_1^{2p+s+1} z_2^{2q+s+1}}
  \overline{\biggl ( {\rm tr} [(V_k)^{2p+s}] \ {\rm tr}
  [(V_k)^{2q+s}] \biggr )_s} \ .  
\label{bin:eq9}
\end{equation}
In equation~(\ref{bin:eq9}), the index $s$ counts the number of
cross--contracted pairs. There are ${2p+s \choose s} {2q+s \choose s}$
different ways of cross--contracting the $V_k$'s, while there are
$(2p-1)!! \, (2q-1)!!$ ways of pairwise contracting the remaining
$V_k$'s.  For the latter, we use the binary correlation
approximation. Thus, each one of these contractions yields a factor
unity, irrespective of the position where the contracted $V_k$'s
appear in each of the traces. The result is
\begin{eqnarray}
\overline{g(z_1) g(z_2)} &=& \frac{1}{N^2} \sum_{s = 0}^{\infty} 
  \sum_{p,q = 0}^{\infty} \frac{(2p-1)!! (2q+1)!!}{z_1^{2p+s+1}z_2^{2q+s+1}} 
  {2p+s \choose s} {2q+s \choose s} \nonumber \\
&& \qquad \qquad \times \,
  \overline{ \biggl( {\rm tr} [(V_k)^{s}] \ {\rm tr}
   [(V_k)^{s}] \biggr )_s } \ .
\label{bin:eq10}
\end{eqnarray}

In reference~\cite{ben01b}, it was argued that in the limit $l\to
\infty$ the terms with $s \neq 0$ become negligibly small in
comparison with the terms with $s = 0$. The latter correspond to the
unlinked contributions. Upon resummation, these terms factorize into
the product $\overline{g(z_1)} \cdot \overline{g(z_2)}$. Therefore,
the two--point correlation function $R_2(z_1,z_2)$ defined in
equation~(\ref{bin:eq2}) vanishes asymptotically in the limit $l
\rightarrow \infty$ and, thus, implies a Poissonian spectrum. In
detail, the argument is the following.

Observing that the binomial factors and the powers of $z_1$ and $z_2$
in equation~(\ref{bin:eq10}) do not depend on $l$ or on $m$, and
recalling that $|z_1| \sim |z_2|$, we see that the only $l$--dependent
factor is
\begin{equation}
T_s= \frac{1}{N^2} \overline{ \biggl( {\rm tr} [(V_k)^{s}] \ 
   {\rm tr} [(V_k)^{s}] \biggr )_s } \ .
\label{bin:eq11}
\end{equation}
Obviously $T_0 = 1$ and, thus, all unlinked contributions are constant
and independent of $l$. Using counting arguments, the dilute limit,
and Stirling's formula, one obtains the asymptotic estimate $T_s \sim
l^{-sk}$ for $T_s$~\cite{ben01b}. This result implies that all
connected contributions vanish in the limit $l\to \infty$ at least as
$l^{-k}$. We note that in the dilute limit, $l^{-k}$ vanishes as a
power of the logarithm of the Hilbert space dimension $N$. Hence the
two--point function $R_2(z_1,z_2)$ vanishes. This line of reasoning
can also be applied to the $n$--point correlation function with $n>2$
showing that the spectrum is Poissonian.

This argument has the obvious flaw that it does not display the role
of $\varepsilon$, the difference of the energy arguments of the two
Green functions. By the same token, the argument does not display the
difference between the connected part of the two--point function
(where $z_1$ and $z_2$ carry infinitesimal imaginary increments of
opposite signs) and the product of two advanced (or two retarded)
Green functions (where $z_1$ and $z_2$ carry infinitesimal imaginary
increments of the same sign). In the case of canonical RMT, we know
that in the first case ($z_1$ and $z_2$ carry infinitesimal imaginary
increments of opposite signs) the analogue of $R_2$ does not vanish
asymptotically (in fact, the asymptotic value of this function
determines the $\Delta_3$--statistic and the stiffness of the
spectrum) while in the second case ($z_1$ and $z_2$ carry
infinitesimal imaginary increments of the same sign) the analogue of
$R_2$ vanishes asymptotically. This implies that in the case of
canonical RMT, the analogue of $R_2$ has a discontinuity across the
real energy axis. It is to be expected that such a discontinuity
exists likewise for $R_2(z_1,z_2)$.

Srednicki~\cite{sre02} has recently drawn attention to this singular
behaviour and has considered the convergence properties of the sums in
equation~(\ref{bin:eq9}) in the light of this
question. Equation~(\ref{bin:eq10}) is symmetric in the indices $p$
and $q$. This suggests writing $R_2(z_1,z_2)$ as
\begin{equation}
R_2(z_1,z_2) = \sum_{s=1}^\infty { g_s(z_1) g_s(z_2) \over 
  \overline{g(z_1)} \cdot \overline{g(z_2)}} T_s \  ,
\label{bin:eq12}
\end{equation}
where the function $g_s(z)$ is given by
\begin{equation}
g_s(z) = \sum_{p=0}^\infty \frac{(2p-1)!!}{z^{2p+s+1}} {2p+s \choose
  s} = \frac{1}{s!\, z^{s+1}} \sum_{p=0}^\infty \frac{(2p+s)!}{p!\, (2
  z^2)^p}\ .
\label{bin:eq13}
\end{equation}
In equation~(\ref{bin:eq12}) we have used the fact that $g_{s=0}(z) =
\overline{g(z)}$, see equation~(\ref{bin:eq3}). Using Borel resummation and
working out the resulting integrals as in the case of the one--point
function, one is led to the integral representation~\cite{sre02}
\begin{equation}
g_s(z^\pm) = \frac{(\mp i)^{s+1}}{s!} \int_0^\infty du \, u^s
  \exp \biggl[-\frac{u^2}{2}\pm i z^{\pm}u \biggr] \ .
\label{bin:eq14}
\end{equation}
The upper plus (minus) sign refers to a positive (negative)
infinitesimal imaginary energy increment, respectively. Using
equation~(\ref{bin:eq14}) in equation~(\ref{bin:eq12}) yields
\begin{equation}
\fl
R(z_1^{+},z_2^{\pm}) = {\mp 1 \over \overline{g(z_1^+)} \cdot
  \overline{g(z_2^\pm)}} \int_0^\infty du\, dv\, e^{-(u^2+v^2)/2}
  e^{i(z_1^{+}u \pm z_2^{\pm}v)} F(\mp u v) \ ,
\label{bin:eq15}
\end{equation}
where the function $F(y)$ is defined as
\begin{equation}
F(y) = \sum_{s=1}^\infty \frac{y^s}{(s!)^2}T_s \ .
\label{bin:eq16}
\end{equation}

Srednicki points out that, in order for both $R_2(z_1^+,z_2^+)$ and
$R_2(z_1^+,z_2^-)$ to be well defined, $F(y)$ must be free of
singularities on the real axis, in which case the integrals can be
performed ``without further (arbitrary) regularization''. Using the
case $k = 1$, he shows that $F(y)$ does have a singularity. This
then invalidates the conclusion drawn in reference~\cite{ben01b} that the
spectrum is Poissonian. Srednicki concludes that the spectral
statistics of the EGE($k$) in the dilute limit remains an unsolved
problem~\cite{sre02}. The example $k = 1$ considered by Srednicki is
all the more remarkable because in this case, it is clear from
independent arguments that the spectrum is Poissonian for $m \gg 1$.

\section{Correlated Matrix Elements: The Limiting Ensembles} 
\label{corr}

Differences between the spectral fluctuations of EGE($k$) and
canonical RMT may be attributed to the fact that in EGE($k$), the
many--body matrix elements are strongly correlated while in RMT,
they are not. The correlations are apparent from the definition of
the ensemble, equation~(\ref{3}): Matrix elements not related by symmetry
which involve different many--body states $\psi_{k,\alpha}$, may
have the same value. It is instructive to study whether and how these
correlations of the matrix elements influence the spectral statistics.
We do so by maintaining the {\it graphical structure} of the EGE($k$).
By this we mean that many--body matrix elements which vanish for the
EGE($k$) continue to have the value zero. We modify only the values
of the non--vanishing matrix elements in such a way as to display the
role of the correlations. It is in this spirit that we construct two
limiting ensembles, one with maximum and one with minimum correlations
among the matrix elements. Both ensembles have the same graphical
structure as the embedded ensembles. The EGE($k$) will be seen to lie
between these two limiting ensembles. For simplicity we consider only
Fermions. We essentially follow reference~\cite{ben01b}.

The graphical structure of the embedded ensembles may be displayed by
assigning to each Hilbert space vector $\psi_{k,\alpha}$ a vertex
$\alpha$, and to each non--diagonal matrix element $\langle \mu | V_k
| \nu \rangle$ which is not identically equal to zero, a link
connecting the vertices $\mu$ and $\nu$. The diagonal matrix elements
$\langle \mu | V_k | \mu \rangle$ are represented by loops attached to
the vertex $\mu $. With the exception of the loops, the resulting
structure is referred to as a ``regular graph'' in the mathematical
literature~\cite{mar71}. (For Bosons, the resulting structure is not a
regular graph.) The number of vertices is obviously given by $N$, the
dimension of Hilbert space. For Fermions, the number $M$ of links
emanating from any given vertex is the same for all vertices and given
by
\begin{equation}
\label{corr:eq1}
M = \sum_{s = 1}^k {m \choose s} {l-m \choose s} \ . 
\end{equation}
This result is obtained by counting the non--diagonal matrix elements
$\langle \mu | V_k | \nu \rangle$ that connect states which differ in
the occupation numbers of at most $k$ particles. For $k < m$, we have
$M < N-1$ while $M = N-1$ for $k = m$. The number $P$ of independent
links is given by the number of matrix elements above the main diagonal
which do not vanish identically. Hence,
\begin{equation}
\label{corr:eq2}
P = \frac{1}{2} M N .
\end{equation}
Obviously, $N$, $M$ and $P$ do not depend on the symmetry parameter
$\beta$.

\begin{figure}
\centerline{\includegraphics[width=11cm,angle=90]{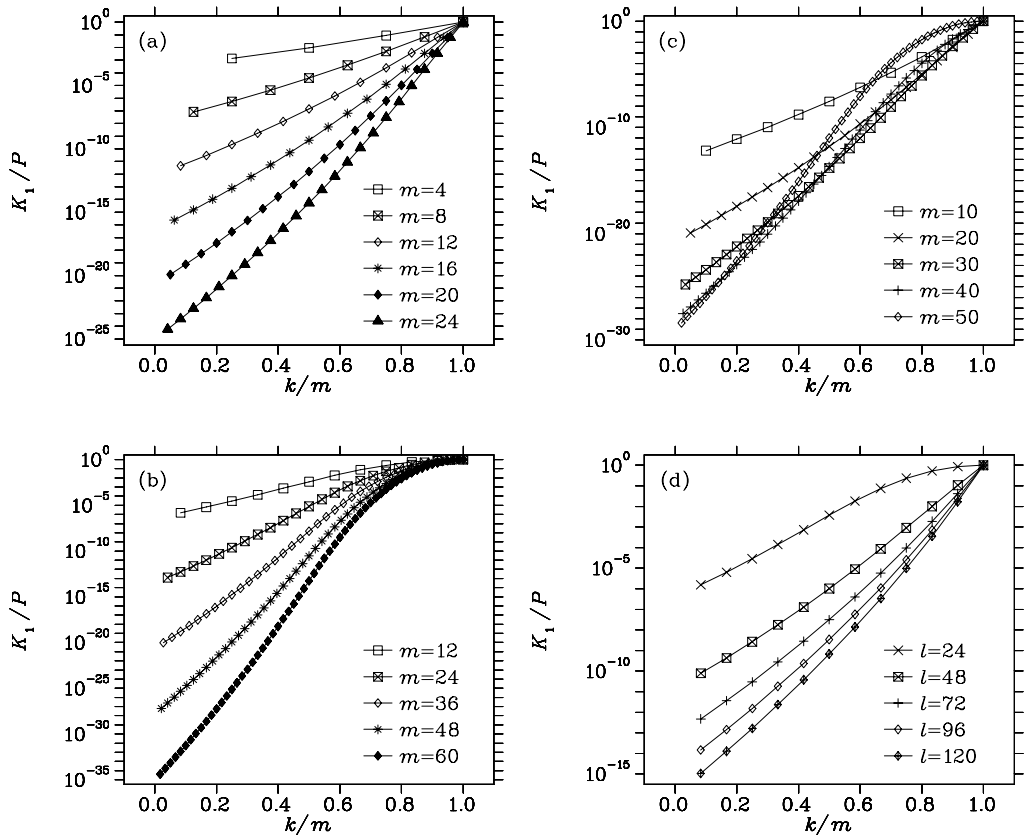}}
\caption{The ratio $K_1/P$, see equations~(\ref{corr:eq2}) and
  (\ref{corr:eq3}), on a logarithmic scale versus $k/m$. Panel (a):
  $f = m/l = 1/5$. Panel (b): $f = m/l = 1/2$. Panel (c): $l = 100$.
  Panel (d): $m = 12$. Taken from reference~\cite{ben01b}.}
\label{f_4}
\end{figure}

To obtain a measure for the correlations between matrix elements, we
define the number $K_{\beta}$ of independent random variables. This
number differs for the unitary and the orthogonal ensemble and is
given by
\begin{equation}
\label{corr:eq3}
K_{\beta} = \frac{\beta}{2} {l \choose k} \biggl[{l \choose k} +
\delta_{\beta 1} \biggr] \ .
\end{equation}
The ratio $K_{\beta}/P$ of the number $K_{\beta}$ of independent
random variables and the number $P$ of different links serves as a
measure of the correlations. For $k \ll m$, $K_{\beta}/P$ is much
smaller than one. It approaches the value $\beta (N + \delta_{\beta
1})/(N - 1)$ monotonically from below as $k$ approaches $m$. This
shows that for $k \ll m$ there are strong correlations between matrix
elements on different links. The correlations disappear as we approach
the canonical limit $k = m$. This is illustrated for several
parameters $k$, $m$ and $l$ in figure~\ref{f_4}.

We now define the two limiting ensembles by assigning a minimum and a
maximum number of independent random variables to the graph structure
of the embedded ensembles. For the EGE($k$), the parameter
$K_{\beta}/P$ lies between the values associated with these limiting
ensembles. Obviously, the minimum number of random variables is
one. The corresponding ensemble EGE$_{\rm min}$($k$) is defined in
terms of the matrix elements $\langle \mu | V_k^{\rm min} | \nu
\rangle$ of the operator
\begin{equation}
\label{corr:eq4}
V_k^{\rm min} = v \sum_{\alpha \gamma}  \psi^{\dagger}_{k,\alpha} 
  \psi_{k,\gamma} \ .
\end{equation}
The factor $v$ is a real (complex) Gaussian random variable for $\beta
= 1$ ($\beta = 2$, respectively). Without loss of generality we may,
however, put $\overline{|v|^2} = 1$, removing the distinction between
the unitary and the orthogonal cases. The ensemble EGE$_{\rm max}(k$)
containing the maximum number of independent random variables is
obtained by assigning, within the constraints imposed by symmetry, a
different random variable $v_{\mu \nu}$ to each link of the graph. The
matrix elements of EGE$_{\rm max}(k$) are given by
\begin{equation}
\label{corr:eq5}
\langle \mu | V_k^{\rm max} | \nu \rangle = v_{\mu \nu} 
  \sum_{\alpha \gamma} \langle \mu | \psi^{\dagger}_{k,\alpha}
  \psi_{k,\gamma} | \nu \rangle \ .
\end{equation}
For $\beta = 1 \ (\beta = 2)$, the matrix $v_{\mu \nu}$ is real
symmetric (complex Hermitean, respectively). Elements not connected by
symmetry are uncorrelated Gaussian random variables with mean value
zero and variance $\overline{v_{\mu \nu} v_{\mu' \nu'}} = \delta_{\mu
  \nu'} \delta_{\nu \mu'} + \delta_{\beta 1} \delta_{\mu \mu'}
\delta_{\nu \nu'}$. 

The ensemble EGE$_{\rm min}(k$) is fully integrable and has spectral
fluctuations which are not of Wigner--Dyson type. For $k=1$, one finds
two different degenerate eigenvalues $\lambda_1 = v l$ and
$\lambda_2 = 0$, with degeneracies $n_1 = {l-1 \choose m-1}$ and $n_2 =
{l-1 \choose m}$. For $k = m$, the matrix representation of EGE$_{\rm
min}(m$) in Hilbert space carries the entry $v$ on every
element. Diagonalization of this matrix is trivial and gives the
eigenvalues $N v$ (non--degenerate) and zero ($(N-1)$--fold
degenerate). Again, the ensemble is fully integrable, and the spectral
fluctuations are not Wigner--Dyson. Using the supersymmetry method it
can be shown that the spectral fluctuation properties of EGE$_{\rm
max}$($k$) coincide with the predictions of RMT. From these facts, it
is apparent that the limiting ensembles cover the extreme cases of a
fully integrable system and a system with Wigner--Dyson spectral
statistics. Figure~\ref{f_4} then suggests that as $k$ increases, the
spectral fluctuations of both EGOE($k$) and EGUE($k$) may undergo a
gradual transition from Poissonian to Wigner--Dyson behaviour.

\section{Conclusions}
\label{con}

We return to the questions raised at the end of Section~\ref{int}. We
recall that both for $k = m$ and in the dilute limit, Fermions and
Bosons behave qualitatively similarly, while the dense limit for
Bosons is a special case.

(i) What is the shape of the spectral density? Among the four
questions formulated in the Introduction, this is the one to which we
have a nearly complete answer. The moments method shows that there is
a gradual transition from semicircular shape for $k = m$ to Gaussian
shape in the dilute limit. This is supported by the binary correlation
approximation which yields a Gaussian shape in the dilute limit. The
eigenvalue expansion of the matrix of second moments has added the
insight that the semicircle prevails as long as $2k > m$, and that the
transition to Gaussian shape sets in at $2k = m$. The special role of
the value $2k = m$ is due to duality. The case of dense Bosons is
special: It is not ergodic.

(ii) What are the spectral fluctuation properties? The supersymmetry
method suggests that the spectral fluctuations are of Wigner--Dyson
type as long as $2k > m$. It is possible that the range of validity of
Wigner--Dyson spectral statistics extends into the domain $2k \leq m$
although here the supersymmetry method yields also non--universal
contributions. In the dilute limit, the situation is not clear. The
straightforward extension of the binary correlation approximation
yields Poissonian statistics for $l \to \infty$ but relies on
manipulations which are mathematically questionable. The eigenvector
expansion of the matrix of second moments yields expressions which
change smoothly with $k$, $m$, $l$. From these expressions one would not
expect a sudden transition from Poissonian statistics (which surely
applies for $k = 1$) to Wigner--Dyson statistics for $k = 2$. The
group--theoretical approach shows that the part of $V_k$ which
transforms under $U(N_m)$ like a multiple of the unit matrix (and,
thus, leads to Poissonian level statistics) is largest for $k \ll m$.
The study of the correlations between many--body matrix elements
provides intuitive insight into the causes responsible for deviations
from Wigner--Dyson statistics. Dense Boson systems are close to
integrable and display no similarity to Wigner--Dyson spectral
statistics.

(iii) Are these properties universal? In analogy to canonical RMT, we
ask here whether the spectral density and the spectral fluctuation
properties hold irrespective of the Gaussian choice for the $k$--body
matrix elements, i.e., the variables $v_{k;\alpha \gamma}$ defined in
Section~\ref{defi}. This is a meaningful question: We recall that in
canonical RMT, ensembles with non--Gaussian weight factors have been
studied. Since for $m = k$, EGE($m$) is identical to canonical RMT,
non--Gaussian ensembles of EGE($k$) type can certainly be defined. The
analytical methods reviewed in this paper all rely heavily on the
Gaussian distribution of the $v_{k;\alpha \gamma}$'s. At this point in
time, no analytical approach is in sight to answer this question, nor
are we aware of any numerical simulations addressing the issue.
Another aspect of universality is addressed in the group--theoretical
analysis of the embedded ensembles. It shows that the embedded
ensembles for Fermions and Bosons are part of a much wider class of
embedded random--matrix ensembles defined in terms of irreducible
representations of some compact Lie group.

(iv) Are the spectra ergodic? Partial answers to this question come
from numerical simulations, from the moments method, and from the
eigenvalue expansion of the matrix of second moments. Ergodicity can
be expected only in the limit of infinite matrix dimension. For $k = m$
ergodicity holds for all observables. In the dilute limit, the spectral
density is ergodic, but non--ergodic contributions vanish very slowly
(with inverse powers of the logarithm of the dimension of the matrices).
This causes difficulties in the analysis of numerically simulated
spectra. For dense Bosons, even the low moments of the spectral density
are not ergodic. It seems that this is the first known case of a
random--matrix model which is not ergodic in the limit of infinite
matrix dimension. We are not aware of any studies addressing ergodicity
of the spectral fluctuations.

This review shows that after many years of work, the determination of
the spectral density and, especially, of the spectral fluctuation
properties of EGE($k$) still poses difficult problems. The difficulties
are due to the fact that EGE($k$) lacks the symmetry properties of the
canonical ensembles of RMT.

\ack
We thank M. Srednicki for communicating his work prior to publication.
LB has profited from discussions with J. Flores, H. Larralde, F. Leyvraz
and T.H. Seligman. LB acknowledges financial support from the DGAPA--UNAM
projects IN--109000 and IN--112200. 

\section*{References}


\end{document}